\documentstyle[aps]{revtex}
\begin{document}
\voffset=-2mm
\title{Comment on ``Superconducting phases in the presence of Coulomb 
interaction: From weak to strong correlations''}
\author{\it J.J. Rodr\'{\i}guez--N\'u\~nez}
\address{Departamento de F\'{\i}sica-CCNE, 
Universidade Federal de Santa Maria,   
97105-900 Santa Maria/RS, 
Brazil. \\e-m: jjrn@ccne.ufsm.br}
\author{\it A.A. Schmidt}
\address{Departamento de Matem\'atica-CCNE, 
Universidade Federal de Santa Maria,   
97105-900 Santa Maria/RS, 
Brazil. \\e-m: alex@lana.ccne.ufsm.br}
\date{\today}
\maketitle
\begin{abstract}
We examine the paper basic equations of T. Doma\'nski and K.I. 
Wysoki\'nski [Phys. Rev. B {\bf 59}, 173 (1999)], who calculated 
the critical superconducting temperature, $T_c$, in function of 
Coulomb correlations for $s$- and $d$--wave order parameter symmetries. 
We argue that in their gap equation the Coulomb repulsion is 
counted twice. Then, we write down the right gap equation and 
solve it by using a normal state one--particle Green function 
which gives a Mott metal--insulator transition. Our numerical 
results for $T_c$ vs $U$ shows that $U$ is detrimental 
to superconductivity as it was found by Doma\'nski and 
Wysoki\'nski.\\
\\
Pacs numbers: 74.20.-Fg, 74.10.-z, 74.60.-w, 74.72.-h
\end{abstract}


\indent In a recent paper\cite{DW}, a study of the evolution of the 
superconducting phases in a model with competing short--range 
attractive ($W$) and on--site ($U$) interactions is presented. The authors 
have evaluated the superconducting critical temperature ($T_c$) at 
the mean field level for the s- and d--wave superconducting order 
parameter symmetries under the influence of correlations ($U$) using 
several approximations ($HF$, second order perturbation theory ($SOPT$) 
and the alloy analogy approximation ($AAA$)). We find that:\\
\indent {\bf 1}- Eq.\ (2) of \cite{DW} counts the Coulomb 
repulsion twice: once 
for the normal state one--particle 
Green function, $G_N(\vec{k},i\omega_n)$, 
and again for the mean field gap equation. 
The right expression should be:
\begin{equation}\label{rightgap}
\Delta_{\vec{k}} = -\frac{T}{N}\sum_{\vec{q},n}
\frac{W_{\vec{q}-\vec{k}}\Delta_{\vec{q}}}{|G_N(\vec{k},i\omega_n)|^2 
+ |\Delta_{\vec{q}}|^2}~~~,
\end{equation}
\noindent where $G_N(\vec{k},i\omega_n)$ is the normal state one--particle 
Green function for which $\Delta_{\vec{q}} \equiv 0$.\\
\indent {\bf 2}- To avoid double counting the Coulomb interaction 
term we use a $G_N(\vec{k},i\omega_n)$ which gives a 
metal--insulator transition ($MIT$). Therefore, we 
propose the following academic well behaved 
$G_N(\vec{k},i\omega_n)$\cite{ILM}:
\begin{equation}\label{hubbardIII}
G_N(\vec{k},i\omega_n) 
= \frac{1-\rho}{i\omega_n + \mu - \varepsilon_{\vec{k}}} + 
\frac{\rho}{i\omega_n + \mu - \varepsilon_{\vec{k}}- U}~~~,
\end{equation} 
\noindent with a critical value of $U$ of $U_c = 2D$ (the bandwidth). With 
the use of Eq.\ (\ref{hubbardIII}), the equations 
for the mean--field superconducting critical temperature, $T_c$, 
and the self--consistent particle density, $\rho$, are
\begin{eqnarray}\label{Tc}
\frac{1}{V} = - \frac{1}{N_s}\sum_{\vec{k}} \left[ 
\frac{(1-\rho)^2\tanh\left(\frac{
\varepsilon_{\vec{k}}-\mu}{2T_c}\right)}{2(\varepsilon_{\vec{k}}-\mu)}
+ \frac{\rho^2\tanh(\frac{
\left(\varepsilon_{\vec{k}}-\mu+U\right)}{2T_c})}
{2(\varepsilon_{\vec{k}}-\mu+U)} +
\frac{\rho(1-\rho)\left[\tanh(\frac{\left(\varepsilon_{\vec{k}}-
\mu+U\right)}{2T_c}) + 
\tanh(\frac{\varepsilon_{\vec{k}}-\mu}{2T_c})\right]}
{(2(\varepsilon_{\vec{k}}-\mu)+U)}\right]~~~. 
\end{eqnarray}
\noindent and
\begin{eqnarray}\label{Tc1}
\rho &=& \frac{1}{2N_s}\sum_{\vec{k}} \left[1 - \left(1-\rho\right)
\tanh\left(\frac{\varepsilon_{\vec{k}}-\mu}{2T_c}\right) - 
\rho \tanh\left(\frac{\varepsilon_{\vec{k}}-\mu+U}{2T_c}\right)\right] 
~~~,
\end{eqnarray}
\noindent from where we recuperate  
$T^{BCS}_c$ when $U/W = 0$. We have chosen a flat free density of states, 
i.e., $N_L(\epsilon) = 1/2D$ for $-D \leq \epsilon 
\leq +D$ and zero otherwise. Let us say that 
Eq.\ (\ref{hubbardIII}) is the {\it exact} solution of the normal 
state Hamiltonian $H_U$ and the perturbation part is the attractive 
interaction between Cooper pairs, $V \neq f(\vec{k})$ in our case. The case 
of $V = f(\vec{k})$ will not be discussed here. Thus, our full 
Hamiltonian can be written as
\begin{equation}\label{twoterms}
H = H_U + H_V~~~.
\end{equation}
\indent From Eq.\ (\ref{twoterms}) we conclude that the 
perturbation is $H_V$, i.e., we can use a mean field analysis to the 
second term of Eq.\ (\ref{twoterms}). In other 
words, one knows the ``exact'' solution of $H_U$ and our problem 
at hand is to calculate the solution of the full Hamiltonian $H$. 
Then, the normal state one--particle 
Green function must be valid for any value of $U$. Because of this, 
Eq.\ (\ref{rightgap}) is fully justified. Under these circunstances, 
it would be an error 
to perform mean field approximation on both $U$ and $W_{\vec{q}}$.\\
\indent In Fig.\ 1 we present $T_c$ vs $U$ for several values of 
$V$. In (a) $\mu = 0.25 (U/2)$; (b) $\mu = 0.50 (U/2)$; (c) 
$\mu = 0.75 (U/2)$ and (d) $\mu = 1.0 (U/2)$. For $\mu = U/2$ we 
are at half--filling. We have chosen $2D = 1.0$. 
From Fig.\ 1 we observe that there is 
critical value of $U$ beyond which $T_c$ is zero. This clearly 
shows that the Coulomb interaction (correlations) conspire against 
superconductivity. These results agree with the ones found in  
Ref.\cite{DW}. Our calculations have been performed 
assuming that the Hartree shift due to Cooper pairing, $\rho V$, is 
the same both in the normal and superconducting phases. This is 
the reason that we have not renormalized the chemical potential with 
the pairing interaction. Our 
approximation (Eq.\ (\ref{hubbardIII})) is an academic one  
because the weights of the spectral 
functions of $G_N(\vec{k},i\omega_n)$ are not 
$\vec{k}$--dependent (they are $\alpha_1(\vec{k}) = 1-\rho$ 
and $\alpha_2(\vec{k}) = \rho$, respectively).  
However, other approximations, besides the ones used in Ref.\cite{DW} 
and our Eq.\ (\ref{hubbardIII}),  
can be employed, for example the one of Nolting\cite{Nolting} 
which gives a metal--insulating transition if the band narrowing 
factor term, $B(\vec{k})$, is properly treated\cite{JJRNetal}. For 
high values of $|V|$ we should take into account the effect of 
superconducting pair fluctuations as it 
has been done by Schmid\cite{albert} and others\cite{Tifrea}.\\
\indent In conclusion we have justified our gap equation (Eq.\ 
(\ref{rightgap}). See also Eq.\ (\ref{twoterms})) which modifies Eq.\ (2) 
of Ref.\cite{DW}. By proposing Eq.\ (\ref{rightgap}) we  avoid  
doubling counting of the Coulomb interaction ($U$). At the same 
time, we have used a normal state one--particle Green function 
(Eq.\ (\ref{hubbardIII})) which yields the metal--insulator transition. 
Also, we have presented results of $T_c$ vs $U$ for several values 
of $V$ and of $\mu = \alpha (U/2)$, where $\alpha = 0.25$; $0.50$; 
$0.75$ and $1.00$. The present results agree with the conclusions 
of Ref.\cite{DW} that correlations conspire against superconductivity.\\
 
\noindent {\bf Acknowledgements}\\
\indent We thank Prof. Sergio Garcia Magalh\~aes for interesting 
discussions. The authors thank partial support from 
FAPERGS--Brasil, CONICIT--Venezuela (Project F-139), CNPq--Brasil.
%

\noindent {\Large Figure Captions}\\

\noindent Figure 1. $T_c$ vs $U$, for different values of 
$V$, i.e., $V = -0.50$; $-1.00$; $-1.50$ and $-2.00$. (a) 
$\mu = 0.25 (U/2)$; (b) $\mu = 0.50 (U/2)$; (c) $\mu = 
0.75 (U/2)$; and (d) $\mu = 1.00 (U/2)$. 
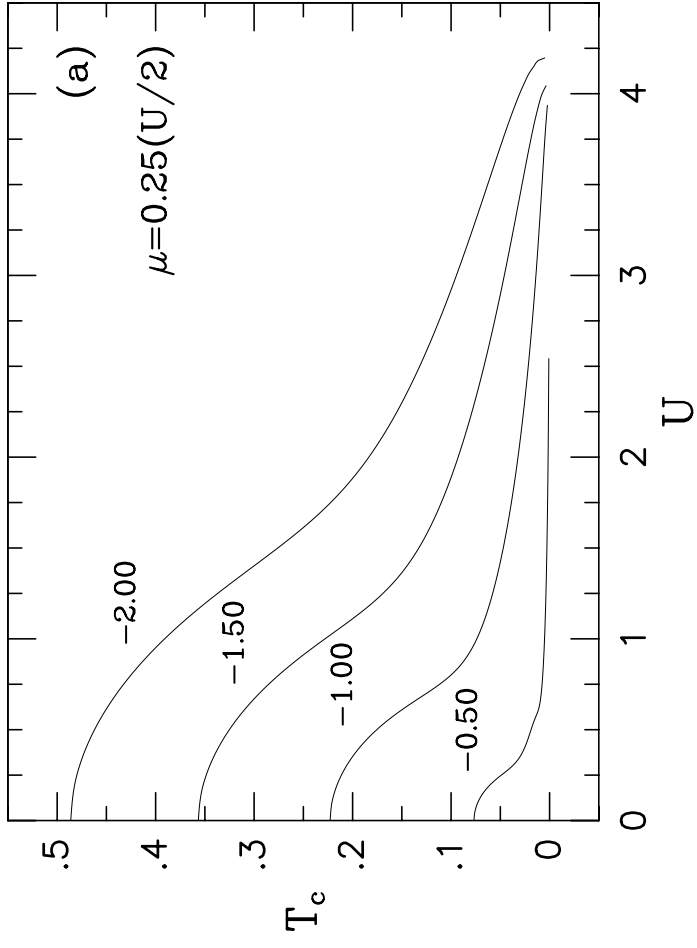
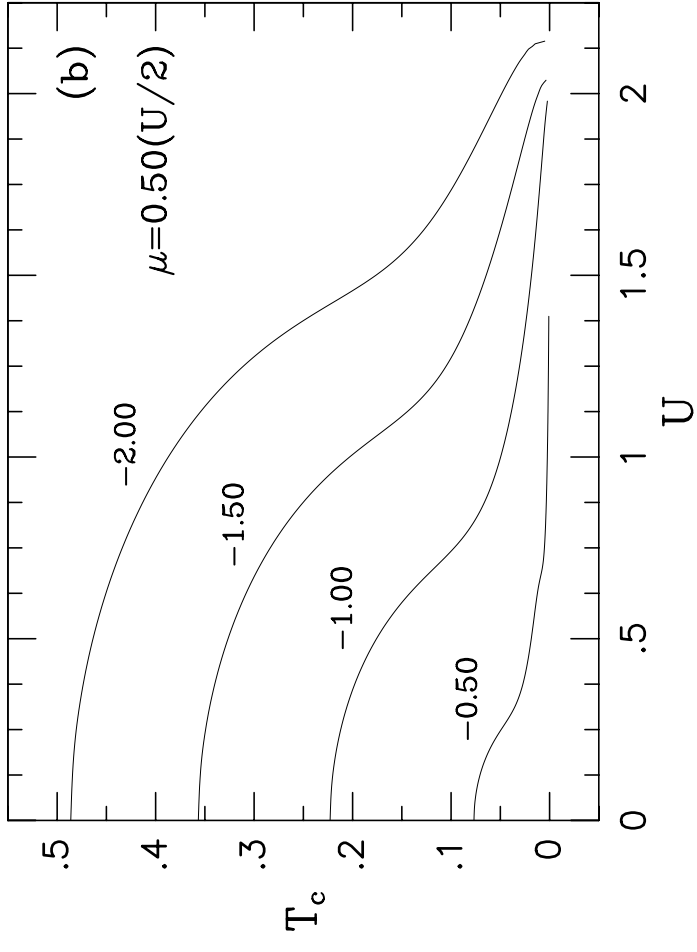
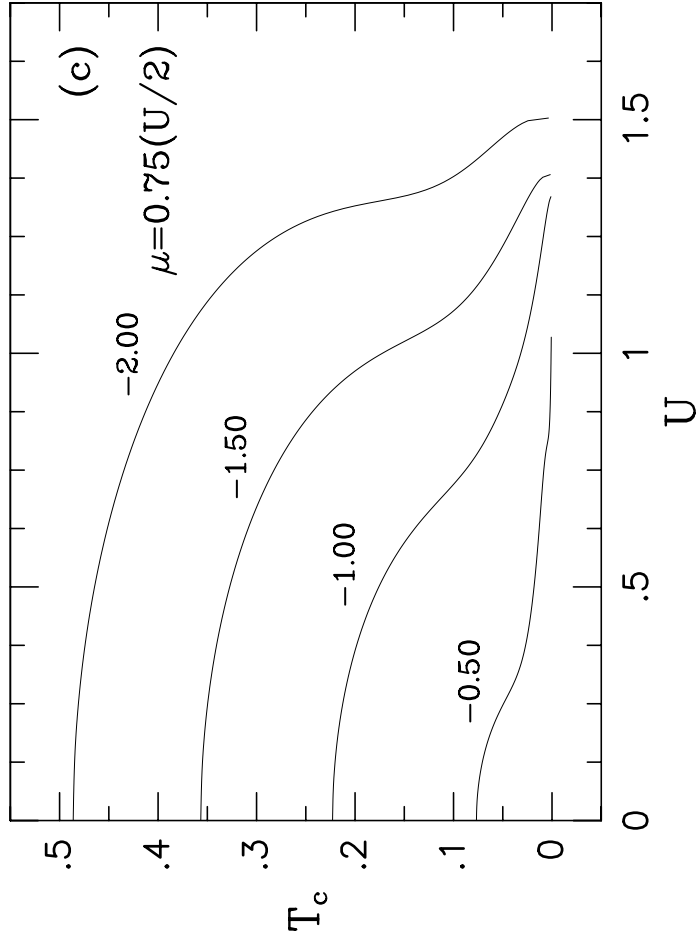
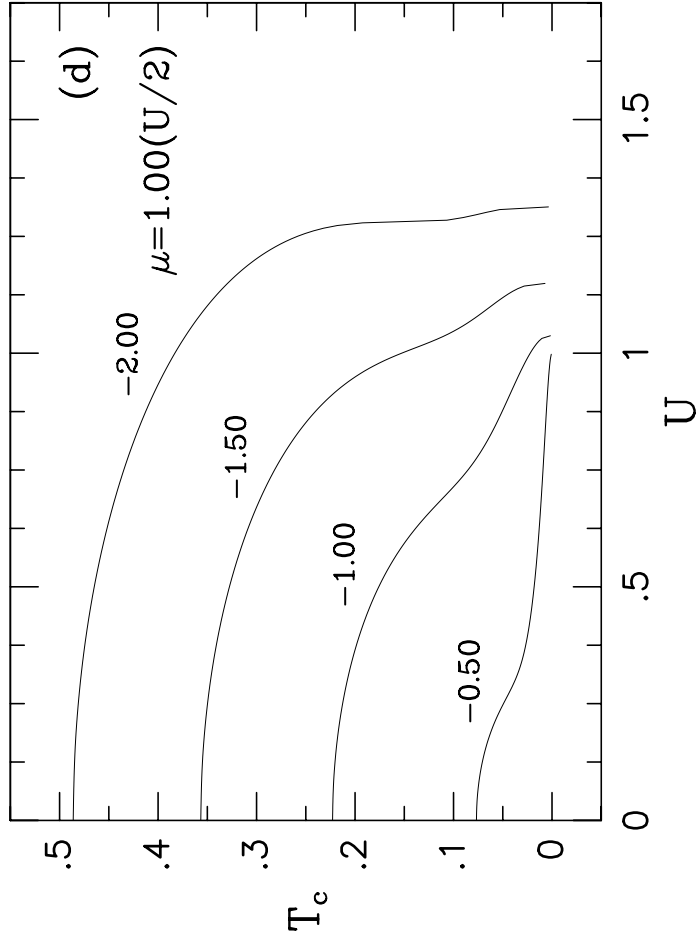
\end{document}